\documentstyle[aps,twocolumn,floats,epsf]{revtex}

\begin{document}

\def\B{\mbox{\bf B}}
\def\x{\mbox{\bf x}}
\def\z{\mbox{\bf z}}
\def\u{\mbox{\bf u}}
\def\j{\mbox{\bf j}}
\def\A{\mbox{\bf A}}
\def\e{\bf {\hat e}}
\def\m{\bf {\hat m}}
\def\s{\bf {\hat s}}
\def\E{\bf {\hat E}}
\def\M{\bf {\hat M}}
\def\S{\bf {\hat S}}
\def\v{\bf v}
\def\zhat{\hat {\bf z}}
\def\psibar{\hat \psi}
\def\V{\mbox{\bf V}}
\def\u{\mbox{\bf u}}
\def\G{\mbox{\bf G}}
\def\A{\mbox{\bf A}}

\draft

\title{Hamiltonian structure of hamiltonian chaos}
\author{X. Z. Tang \footnote{Email: tang@chaos.ap.columbia.edu}}
\address{Department of Applied Mathematics and Statistics,
The University at Stony Brook, Stony Brook, NY 11794\\
and Department of Applied Physics, Columbia University, New York, NY 10027} 
\author{A. H. Boozer}
\address{Department of Applied Physics, Columbia University, New York,
NY, 10027\\
and Max-Planck Institut f{\"u}r Plasmaphysik, Garching, Germany}

\date{\today}
\maketitle

\begin{abstract}
From a kinematical point of view, the geometrical 
information of hamiltonian chaos is given by the (un)stable directions,
while the dynamical information is given by the Lyapunov exponents.
The finite time Lyapunov exponents are of particular
importance in physics. 
The spatial variations of the finite time Lyapunov exponent
and its associated (un)stable direction are related. 
Both of them are found to be determined by a new hamiltonian of
same number of degrees of freedom as the original one.
This new hamiltonian defines a flow field with characteristically
chaotic trajectories. The direction and 
the magnitude of the phase flow field 
give the (un)stable direction and the finite time Lyapunov exponent 
of the original hamiltonian.
Our analysis was based on a $1{1\over 2}$ degree
of freedom hamiltonian system.
\end{abstract}


\section{Introduction}
\label{section:introduction}

Many systems that are chaotic such as the stochastic magnetic field
line in toroidal plasma confinement devices 
have a hamiltonian representation \cite{boozer_83,cary}.
The KAM theorem \cite{arnold} deals with the 
onset of chaos in hamiltonian systems,
and the description of the kinematics of developed chaos 
involves the ergodic theorem \cite{ruelle}.
Hamiltonian dynamics naturally give rise to a differentiable dynamical 
system where the multiplicative ergodic theorem
of Oseledec is applicable \cite{oseledec}.
Oseledec's theorem gives two aspects of chaos.
First the sensitivity of the dependence on initial conditions
is measured by various infinite time Lyapunov exponents, 
{\it i.e.}, the dynamical information of chaos.
Secondly the characteristic directions associated
with these Lyapunov exponents give
the geometrical aspect of chaos.
That is: if two points are separated along different directions
at the initial time, they can diverge or converge exponentially 
at different characteristic Lyapunov exponents.

Ergodic theory treats the time asymptotic limit, in which the infinite
time Lyapunov exponents are constants and the characteristic
directions are functions of position only.
For finite time, there is a convergence issue for both the Lyapunov
exponents and their associated characteristic directions.
The convergence of the characteristic directions is exponential, so the 
geometrical aspect of chaos at finite time is well described by its time 
asymptotic limit. This is not the case for Lyapunov exponents.
Finite time Lyapunov exponents suffer from a notoriously slow 
convergence problem \cite{slow_convergence}. 
Its spatial and temporal dependence was discussed by 
Tang and Boozer \cite{tang}, 
who gave a direct
link between the convergence function in finite time Lyapunov exponent
and the geometry of the vector field defined by the corresponding
characteristic direction.
It should be emphasized that
most applications of practical interest involve a finite duration of time.
Hence the finite time properties of chaos, rather than the asymptotic 
properties of chaos, are of real concern.

In \cite{tang} we found 
that the spatial variation of the finite time Lyapunov
exponent and the corresponding characteristic direction are not 
independent of each other. The exact relation was shown 
in \cite{tang} and is restated
later in this paper, 
equations (\ref{exponent_1},\ref{exponent_2},\ref{exponent_3}). 
Besides its practical importance such 
as those in transport studies \cite{tang},
it also leads to a conceptual advance in the understanding of chaos. 
The finite time Lyapunov exponent (e.g. $\lambda$) and its
characteristic direction (e.g. ${\s}_\infty$) 
can be described by another 
hamiltonian in the same phase space with the same number of degrees of 
freedom as the original hamiltonian.
If one constructs the corresponding vector field (necessarily divergence-free)
associated with the phase space trajectory of this new hamiltonian,
the magnitude of this vector field gives the local finite time Lyapunov 
exponent while the direction of the vector field gives the corresponding
characteristic direction.
The new hamiltonian is also chaotic in the same region as the parent
hamiltonian. One could characterize the chaos in this new hamiltonian
by invoking a third generation hamiltonian. 
Consequently, a hierarchy of hamiltonian can be bootstrapped from
the original hamiltonian of a conservative system that is chaotic.   

\section{Globally divergence-free fields and hamiltonian mechanics}
\label{section:hamiltonian}

A large class of conservative systems which exhibit chaotic
behavior has a hamiltonian representation. Two of the well
known examples are the magnetic field ${\B}$ and the velocity field 
${\v}$ of a divergence-free fluid \cite{boozer_enc}. 
An arbitrary divergence-free vector
${\G}(x,y,z)$ can be written in the so-called canonical form,
\begin{equation}
\label{canonical}
{\G} = \nabla\psi\times\nabla\theta + \nabla\phi\times\nabla\chi.
\end{equation}
The function $\chi(\psi,\theta,\phi)$ is the hamiltonian of the
${\G}$ field lines.
The ${\G}$ field line is the trajectory ${\x}(\tau)$ given by 
equation
\begin{equation}
\label{field_line}
{d{\x}\over {d\tau}} = {\G}({\x}).
\end{equation}
To find the ${\G}$ field line in $\psi,\theta,\phi$ coordinates,
one has to invert the transformation equation ${\x}(\psi,\theta,\phi).$
Combining the transformation equation ${\x}(\psi,\theta,\phi)$ and
the field line trajectory, equation (\ref{field_line}),
one arrives at the familiar Hamilton's equations,
\begin{equation}
{d\psi\over{d\phi}} = - {\partial\chi\over{\partial\theta}},\,\,\,
{d\theta\over{d\phi}} = {\partial\chi\over{\partial\psi}},
\end{equation}
where $\chi$ is the hamiltonian, $\theta$ the canonical position,
$\psi$ the canonical momentum, and $\phi$ the canonical time
\footnote{In this extended `phase' space, 
there are three independent coordinates
and hence three Lyapunov exponents.}.
Since continuous transformation, such as ${\x}(\psi,\theta,\phi),$
preserve topological properties, questions regarding the integrability 
of the field line are then answered by the hamiltonian 
$\chi(\psi,\theta,\phi)$ alone.

It should be noted that only a {\it globely} divergence-free field
can be represented by the form of equation (\ref{canonical}), a
requirement set by Poincare's lemma \cite{schutz}. 
Furthermore, the divergence-free
field ${\G}$ should not vanish in the region of interest.

A hamiltonian representation of such a vector field is desirable
since an array of well developed techniques from hamiltonian mechanics
are available. 
For example, the hamiltonian description of a magnetic field
has played a major role in the theory of toroidal plasmas \cite{boozer_enc}.  
 
Before we proceed to the next section which is on the ergodic theorem,
a digression on the applicability of the multiplicative 
ergodic theorem to hamiltonian chaos is useful.
A hamiltonian flow, even when it is chaotic, can not support
a nonuniform invariant density on its ergodic subcomponent in phase space. 
However, numerical simulations
appear to offer a contradictive picture: there are concentrations,
or spikes, of the phase space density near the boundary of 
chaotic zones \cite{smith}.
One might question the ergodicity of the chaotic sea 
(the irregular component \cite{meiss})  and
hence the applicability of Oseledec's theorem.
A recent study by Meiss in the case of the standard map \cite{meiss}
appears to reconcile this conflict.
Meiss found that these density spikes are transient and they 
can be explained by a Markov tree model taking into account 
the effects of islands-around-islands structure. 
Hence earlier numerical findings \cite{smith} are not in contradiction 
to an invariant measure, and the phase space density
can be uniform on an irregular component of hamiltonian systems.
Despite the lack of a rigorous theory \cite{meiss}, 
these new numerical evidence
and the simple Markov transport model put forwarded by Meiss should
boost our confidence in the ergodicity of hamiltonian chaos.   

\section{Multiplicative ergodic theorem in Lagrangian coordinates}
\label{section:ergodic}

The multiplicative ergodic theorem of Oseledec \cite{oseledec} complements
KAM theorem \cite{arnold} in understanding hamiltonian chaos,
especially the limiting case in which
the regular components occupy a small portion of the phase space.
This theorem can be understood in terms of a general coordinate
transformation \cite{schutz} between ordinary space ${\x}$ and the Lagrangian 
coordinates ${\xi},$ a widely used coordinate system in fluid mechanics.
A point with Lagrangian coordinates ${\xi}$ is related to ${\x}$
in ordinary space by the integral curve of 
$d{\x}/dt = {\G}({\x})$ from initial time $t_0$ to later time $t$
with the initial condition ${\x}({\xi},t_0) = {\xi}.$
For a pair of fixed $(t_0,t),$ there is a one-to-one mapping between ${\xi}$
and ${\x}({\xi},t).$ There is freedom in the choice of Lagrangian
coordinates ${\xi}$ due to the arbitrariness of $t_0,$
a point that we will revisit in section \ref{section:core}.  
Of course, $t_0$ should be fixed for any chosen set of 
Lagrangian coordinates. $t_0$ is usually set to zero for convenience
of bookkeeping.

In addition to a set of coordinates, one needs a metric to specify
the physical distance between two neighboring points.
Let ${\x}$ be ordinary Cartesian coordinates, then its metric is
a unit matrix, $dl^2 = d{\x}\cdot d{\x}= dx^idx^i.$
The same differential distance can also be specified in
Lagrangian coordinates, $dl^2 = g_{ij} d\xi^id\xi^j,$
where $g_{ij}\equiv {\partial{\x}/\partial\xi^i}\cdot
{\partial{\x}/\partial\xi^j}$ is the metric tensor of the Lagrangian
coordinates $\xi.$
Since ${\x}(\xi,0) = \xi,$ one could also understand
$(d{\x})^2= g_{ij} d\xi^i d\xi^j$
as an equation which relates the initial separation $d\xi^i d\xi^j$
between two neighboring points to their later separation $(d{\x})^2.$
Once this is understood, the interpretation of the metric tensor
$g_{ij}$ in Oseledec's multiplicative ergodic theorem becomes transparent,
{\it i.e.},
$g_{ij}$ is the Oseledec matrix $\Lambda_{ij}$ (for its definition, see 
\cite{oseledec} or \cite{ruelle}).

Instead of taking the limit of
$\lim_{t\rightarrow\infty} (\Lambda_{ij})^{1/2t} = \Lambda$
and diagonalizing the matrix $\Lambda,$ we diagonalize the matrix
$g_{ij}$ first and then take the limit of $t\longrightarrow\infty.$
Since $g_{ij}$ is a positive definite and symmetric matrix, it can
be diagonalized with positive eigenvalues and real eigenvectors,
$$
g_{ij} = \Lambda_l {\e}{\e} + \Lambda_m {\m}{\m} + \Lambda_s {\s}{\s},
$$
with $\Lambda_l\ge \Lambda_m\ge \Lambda_s > 0.$
There are three Lyapunov characteristic exponents associated with
vector field ${\G},$
$$
\lambda_l^\infty=\lim_{t\rightarrow\infty} {\ln\Lambda_l\over{2t}},\,\,\,
\lambda_m^\infty=\lim_{t\rightarrow\infty} {\ln\Lambda_m\over{2t}},\,\,\,
\lambda_s^\infty=\lim_{t\rightarrow\infty} {\ln\Lambda_s\over{2t}}.
$$
For a divergence-free field with a hamiltonian representation, 
$\lambda_l^\infty=-\lambda_s^\infty>0$
and $\lambda_m^\infty=0.$
These are usually called infinite time Lyapunov exponents and
they are constants in one ergodic region.
The eigenvectors ${\e}, {\m}, {\s}$ have well-defined time-asymptotic
limits as well, but they are position dependent,
${\e}_\infty(\xi), {\m}_\infty(\xi), {\s}_\infty(\xi).$
There is a stable manifold theorem in the vicinity of an arbitrary
point $\xi$ \cite{ruelle}. The ${\s}_\infty(\xi)$ is the tangent vector of the 
local stable manifold associated with the Lyapunov characteristic exponent
$\lambda_s^\infty,$ the so-called stable direction.
There is also an unstable direction which is tangent to the local
unstable manifold. The unstable direction 
is defined using the time-reversed dynamics,
$$
{d{\x}\over {dt}} = - {\G}({\x}),
$$
in which $dl^2 = (d{\x})^2 = g_{ij}^{(-)} d\xi^i d\xi^j$
and 
$g_{ij}^{(-)} = \Lambda_l^{(-)} {\e}^{(-)} {\e}^{(-)}
+ \Lambda_m^{(-)} {\m}^{(-)} {\m}^{(-)}
+ \Lambda_s^{(-)} {\s}^{(-)} {\s}^{(-)}.$
The local stable manifold for the backward time dynamics defines
the local unstable manifold for the forward time dynamics.
Two linearly independent vectors define a plane. The intersection
of two planes gives rise to another vector.
One can use the intersection of plane $({\s}_\infty,{\m}_\infty)$
and plane $({\s}_\infty^{(-)}, {\m}_\infty^{(-)})$ to construct
another vector ${\bar {\bf m}}.$
For the case we are interested in this paper, $\lambda_m^\infty=0$
and ${\bar {\bf m}}$ is tangent to the
center manifold. In the general case $\lambda_m^\infty$ could be nonzero,
${\bar {\bf m}}$ is then the tangent vector
of another local (un)stable manifold, corresponding to a (positive) 
negative $\lambda_m^\infty.$
In either case, these three vectors ${\s}_\infty, {\s}_\infty^{(-)},$
and ${\bar {\bf m}}$ span the three dimensional space.
 
The negative Lyapunov exponent means neighboring points converge
exponentially in time, while a positive Lyapunov exponent means
neighboring points diverge exponentially in time.
The ones of most importance are the most negative and the most positive
Lyapunov exponents. 
In applications, one is concerned
with the finite time Lyapunov exponents.
In the case studied in this paper, there are two of them and they
are defined as
\begin{eqnarray}
\lambda_s & \equiv & [\ln\Lambda_s]/{2t} \\
\lambda_s^{(-)} & \equiv & - [\ln\Lambda_s^{(-)}]/{2t}.
\end{eqnarray}    
Obviously $\lambda_s$ is the negative Lyapunov exponent 
and $\lambda_s^{(-)}$ is the positive Lyapunov exponent
for forward time.
It is easy to see that 
any theory for ${\s}_\infty$ and $\lambda_s$ would be applicable to
${\s}_\infty^{(-)}$ and $\lambda^{(-)},$
since they are defined for the same flow except for a reversal of 
the direction of time.
From now on we concentrate on the set of $(\lambda_s, {\s}_\infty)$
and let
\begin{equation}
\label{redefine}
\lambda(\xi,t) \equiv - \lambda_s(\xi,t)
\end{equation}
for convenience in notation.

\section{The hamiltonian nature of the theory of finite time Lyapunov
exponents}
\label{section:core}

Before we proceed, it is useful to summarize the main results of 
ergodic theorem.
Chaos means sensitive dependence on initial conditions.
There are two aspects of chaos which are captured by the 
multiplicative ergodic theorem of Oseledec.
The first one is the dynamical aspect, {\it i.e.}, how sensitive
is the dependence on initial conditions?
This is answered by the Lyapunov characteristic exponents.
The second one is on geometry. Along different directions,
neighboring points do behave differently, {\it i.e.},
different Lyapunov exponent corresponds to different characteristic
directions.
As signified by its name, ergodic theorem treats the time-asymptotic
limit or long time average.
Hence the Lyapunov exponents in ergodic theorem are also called
infinite time Lyapunov exponents which are constants in a chaotic
sea. The geometrical information that is given by the ergodic theorem
is always local, {\it i.e.}, ${\s}_\infty(\xi)$ is a function
of position.
The smoothness of ${\G}({\x})$ implies the smoothness of
${\s}_\infty(\xi)$ as a vector field.
The field line of ${\s}_\infty(\xi)$ is of 
great importance in transport of advection-diffusion type.
For example, the rapid diffusive relaxation of an externally
imposed scalar or vector field occurs only 
along the ${\s}$ lines \cite{tang}.

In applications where finite time is of concern, one needs to
understand the properties of the {\it finite time} Lyapunov exponents.
The finite time Lyapunov exponent has both a time and space dependence.
It is also called local Lyapunov exponent for that reason.
Since, for example, both $\lambda(\xi,t)$ and ${\s}_\infty(\xi)$
are local, one might feel that there could be some relationship which
relates the spatial variation of these two.
This was addressed by our work 
on finite time Lyapunov exponent \cite{tang}. 
We found that the 
finite time Lyapunov exponent $\lambda(\xi,t)$
can always be asymptotically decomposed into 
three main parts with the addition of an exponentially small fourth
(correction) term
\begin{equation}
\label{exponent_1}
\lambda(\xi,t) = \tilde{\lambda}(\xi)/t + 
f(\xi,t)/\sqrt{t} + \lambda^\infty +  O(\exp[-2\lambda(\xi,t) t]),
\end{equation}
where 
\begin{eqnarray}
{\s}_\infty\cdot\nabla_0 f(\xi,t) & = & 0, \label{exponent_2}\\
{\bf {\hat s}}_\infty\cdot\nabla_0\tilde{\lambda}(\xi) 
+ \nabla_0\cdot{\bf {\hat s}}_\infty & = & 0,\label{exponent_3}
\end{eqnarray}
and $\lambda^\infty$ is the infinite time Lyapunov exponent.
We note that $\tilde{\lambda}(\xi)$ is a smooth function of position due to
the smoothness of vector field ${\s}_\infty.$

The correction to the asymptotic decomposition, {\it i.e.},
the fourth term in equation (\ref{exponent_1}), 
becomes exponentially small as $t$ becomes large.
The rate of
the exponential decay is given by the magnitude of 
the local Lyapunov exponent.
Hence the correction term becomes negligible on a time scale of
a few local Lyapunov time.
It should be emphasized that what we presented here
is  an asymptotic form
of a local expression (finite time Lyapunov exponent is an explicit function
of position and time). 
In different regions of a chaotic component, the local Lyapunov exponent
can vary significantly. For example, the stochastic layer would have a much
smaller finite time Lyapunov exponent and hence a longer period during
which the exponentially small correction is still important. However 
that time length is fixed on the order of a few local Lyapunov time, 
which is the characteristic time scale over which the 
stochasticity of a chaotic trajectory starting from a particular 
location is of practical importance.
This should be contrasted with a statistical description of the
finite time Lyapunov exponent, whose practical applicability would be 
affected by the 
extremely long transients for hamiltonian systems to reach to
an invariant distribution by following one chaotic trajectory\cite{meiss}. 

Because $f(\xi,t)$ does not vary along the ${\bf {\hat s}}$ direction, 
thus to exponential accuracy
the variation of the finite time Lyapunov exponent along the stable
manifold is determined by the geometry of the stable manifold alone, 
equation (\ref{exponent_3}).
Notice that everything is local, i.e. functions of position. In particular,
$\tilde{\lambda}(\xi)$ and ${\bf {\hat s}}_\infty(\xi)$ 
do not have a time dependence.
That is: they represent time asymptotic structures. They are of great practical
importance because the equation (\ref{exponent_3}) accurately
describes the spatial variation of the finite time Lyapunov exponent 
along the stable manifold on a time scale of a few local Lyapunov time.
The effect of geometry on a dynamical quantity like the finite time 
Lyapunov exponent is captured by the function $\tilde{\lambda}(\xi)$
alone,
a function that is completely determined by the ${\G}({\x})$ field.

The main points of Tang and Boozer's work on finite time Lyapunov 
exponents \cite{tang} are the  
function $\tilde{\lambda}$ and its relationship 
with ${\s}_\infty.$ These results actually find their roots in
hamiltonian dynamics.
To see that, one can construct a new vector field 
$$
{\bf S}(\xi) \equiv e^{\tilde{\lambda}(\xi)} {\s}_\infty(\xi).
$$
Obviously
${\bf S}(\xi)$ does not vanish anywhere  and 
is globally defined in a chaotic
region, or using Meiss's term, on an ergodic irregular component.
More importantly, ${\bf S}$ is divergence-free 
because of equation (\ref{exponent_3}).
All the necessary information on chaos in field ${\G}({\x})$
associated with negative Lyapunov exponent for forward time, are
contained in this new vector field ${\bf S}.$

On a regular component in phase space, i.e. KAM surface, ${\bf S}$ 
is also  well defined and divergence-free.
The property of a trajectory on a KAM surface is determined by the 
rotational transform $\iota(\psi).$ In explicit form, the trajectory
follows
\begin{equation}
\label{trajectory}
\psi=\psi_0; \,\,\,\,\,
\phi=\phi_0 + \nu_0(\psi) t; \,\,\,\,
\theta=\theta_0 + \iota(\psi) \nu_0(\psi) t
\end{equation}
with $\nu_0(\psi)$ the Jacobian of $(\psi,\phi,\theta)$ coordinates.
The form of equation (\ref{trajectory}) is generic for integrable
trajectories \cite{arnold_new}. An explicit construction of the proper
coordinate system which gives rise to equation (\ref{trajectory})
can be found in \cite{tang_thesis}.  
The ${\s}_\infty$ introduced earlier now takes the form
\begin{equation}
\label{s_kam}
{\s}_\infty \propto (0, \nu_0', \iota'\nu_0+\iota\nu_0').
\end{equation}
where prime denotes a derivative with respect to $\psi,$ the action.
A detailed derivation of equation (\ref{s_kam}) can be found in \cite{tang_thesis}. 
Vanishing $\psi$ component of ${\s}_\infty$ means that
${\s}_\infty$ vector is tangent to the KAM surface.
By definition, two neighboring points along ${\s}_\infty$
direction will converge, but {\it quadratically} in time, which should be
contrasted to an {\it exponential} rate in a chaotic region.
The ${\bf S}$ vector can be simply given by
$$
{\bf S} = ( 0, \nu_0, {(\iota\nu_0)'\over{\nu_0'}}\nu_0).
$$
This would correspond to an integrable trajectory with rotational transform
${(\iota\nu_0)'/\nu_0'}$ on the constant action $\psi_0$ surface.
In other words, if one writes the original field as
$$
{\G} = \nabla\psi\times\nabla\theta + \nabla\phi\times\nabla\chi
$$
and on the constant $\psi_0$ surface
$$
\iota(\psi_0) = \partial\chi/\partial\psi |_{\psi=\psi_0},
$$ 
then the ${\S}$ field will be
$$
{\S} = \nabla\psi\times\nabla\theta + \nabla\phi\times\nabla\tilde{\chi}
$$
and the  hamiltonian $\tilde{\chi}$ for the field line of ${\S}$ satisfies
$$
\partial\tilde{\chi}/\partial\psi |_{\psi=\psi_0}
= {(\iota\nu_0)'/\nu_0'}|_{\psi=\psi_0}.
$$
Cautions should be taken regarding the derivative with respect
to $\psi.$ The surviving KAM tori are parameterized on a fractal
set of action $\psi$ in a perturbed hamiltonian system.
The proper definition of derivative with respect to $\psi$ on the 
surviving KAM tori
invokes Whitney's notion, 
which was discussed in detail by P{\"o}schel \cite{poschel}. 

It should be noted that so far we have established divergence-free
${\S}$ field separately in chaotic region and KAM region.
To achieve divergence-free globally, it actually only requires the 
smoothness of the direction of ${\S}$ fields when crossing the last KAM
surface. A jump in the magnitude of ${\S}$ is allowed since its gradient
is perpendicular to ${\S}$ 
($\nabla\cdot f{\s}_\infty = \nabla f\cdot {\s}_\infty + f \nabla\cdot{\s}_\infty$).
Although an analytical proof has not been found, numerical results
have been obtained to support the continuity of the direction of ${\s}_\infty$
at the border between order and chaos.   

As pointed out in section \ref{section:hamiltonian}, 
a globally divergence-free field admits
a hamiltonian representation. Hence the field line of ${\bf S}$ 
(and ${\s}$) is described by a hamiltonian $\chi_1(\psi_1,\theta_1,\phi_1)$ 
with $\theta_1$ the canonical position, $\psi_1$ the canonical
momentum, and $\phi_1$ the canonical time.
In other words, the characteristics of chaos in the hamiltonian $\chi$ are
now contained in a new hamiltonian $\chi_1.$
$\chi_1$ has the same degrees of freedom as $\chi$ and the field lines of
${\bf S}$ are chaotic just as those of ${\G}.$
Similarly, one could represent chaos in ${\S}\, (\chi_1)$ by another
hamiltonian $\chi_2$ associated with another divergence-free vector
field ${\bf S}'.$ 
Henceforth a hierarchy of hamiltonians is constructed for describing 
the chaos in ${\G}.$

One might be concerned as to the proper counting of the 
degrees of freedom if
${\G}$ is time dependent. In actuality, ${\bf S}$ becomes time
dependent through ${\s}_\infty.$ This comes from the fact that
the specification of Lagrangian coordinates depends on the choice
of initial time $t_0,$ as discussed in last section.
For a time-dependent field ${\G},$ the vector ${\s}_\infty$ is
a function of position and time ${\s}_\infty(\xi,t_0).$
If ${\G}({\x},t)$ is periodic in time $t,$ ${\s}_\infty(\xi,t_0)$ will be
periodic in time $t_0.$ The hamiltonian $\chi_1$ has the exactly
same degrees of freedom as $\chi.$
This small subtlety holds the key to a correct understanding 
of a time dependent field in two dimensions ${\G}(x,y,t)$
which has chaotic field lines.
The time dependence in ${\s}_\infty(\xi_x,\xi_y,t_0)$ assures
that the corresponding ${\bf S}$ field has a time dependence 
and the hamiltonian $\chi_1$ for ${\bf S}$ is of one and a half
degrees of freedom. Hence chaos is allowed for the field line
of ${\s}_\infty(\xi_x,\xi_y,t_0)$ and one ${\s}$ line
fills the entire irregular component of ${\G}.$

As stated earlier, there is a symmetry between
$({\s}_\infty, \lambda_s)$ and $({\s}_\infty^{(-)}, \lambda_s^{(-)}).$
Hence one could construct a set of hamiltonian to describe the 
chaos associated with the positive Lyapunov exponent and
unstable direction for the forward time dynamics, in analogy
to what we have done for the negative Lyapunov exponent 
and stable direction.
 
\section{Summary}
\label{section:remark}

In this paper we study the interrelations of divergence-free field,
hamiltonian dynamics, multiplicative ergodic theorem in hamiltonian
chaos, general coordinate transformation, 
and the theory of finite time Lyapunov exponents.
We argue that finite time Lyapunov exponents and its associated 
characteristic directions are the most important information on chaos
for the purpose of physical applications.
Unlike the infinite time Lyapunov exponents, 
spatial variation of the finite time Lyapunov exponent is directly related
to the geometry of its corresponding characteristic direction.
Both are found to be given by one single new hamiltonian of same degrees
of freedom.
The magnitude of the phase flow field of the new hamiltonian determines
the finite time Lyapunov exponent, while the direction of the phase
flow gives the corresponding characteristic direction.  
The new hamiltonian is chaotic on the irregular component of the original
hamiltonian.  
We hope the point of view presented here 
could stimulate new insights into hamiltonian chaos.

\acknowledgements

One of the authors (Tang) was supported by a NSF University-Industry
Postdoctoral fellowship in Mathematical Sciences
through SUNY Stony Brook.

\end{document}